\documentclass[graybox]{utils/svmult}

\usepackage{type1cm}        
\usepackage{makeidx}         
\usepackage{graphicx}        
\usepackage{multicol}        
\usepackage[bottom]{footmisc}
\usepackage{xcolor}
\usepackage{array}

\usepackage{draftwatermark}
\SetWatermarkText{Preprint}

\SetWatermarkScale{0.6} 
\SetWatermarkColor[gray]{0.9} 
\SetWatermarkAngle{45} 

\usepackage{newtxtext}       %
\usepackage[varvw]{newtxmath}       

\usepackage[colorlinks=true,allcolors=blue]{hyperref}

\newcommand{\changes}[1]{{\color{black}#1}}

\makeindex             

\begin{document}

\title*{Neuromorphic Spintronics}

\author{Atreya Majumdar and Karin Everschor-Sitte}
\institute{Atreya Majumdar \at University of Duisburg-Essen, Lotharstraße 1 47057 Duisburg, Germany \email{atreya.majumdar@uni-due.de}
\and Karin Everschor-Sitte \at University of Duisburg-Essen, Lotharstraße 1 47057 Duisburg, Germany \email{karin.everschor-sitte@uni-due.de}
\and This is not the final version of the chapter. For the final version, please go to the Springer book titled ``Artificial Intelligence and Intelligent Matter".}

\maketitle

\abstract{Neuromorphic spintronics combines two advanced fields in technology, neuromorphic computing and spintronics, to create brain-inspired, efficient computing systems that leverage the unique properties of the electron's spin. In this book chapter, we first introduce both fields—neuromorphic computing and spintronics—and then make a case for neuromorphic spintronics. We discuss concrete examples of neuromorphic spintronics, including computing based on fluctuations, artificial neural networks, and reservoir computing, highlighting their potential to revolutionize computational efficiency and functionality.}

\section{Introduction}
\label{sec:1}

The invention of computers has been instrumental in ushering in a new era where we can quickly perform large computations in the blink of an eye and store vast amounts of data. This advancement can be directly attributed to the development of electronics, which has enabled efficient miniaturization and high processing speeds. The development of electronic technology has been so rapid that the number of transistors in integrated circuits has doubled every two years since 1970. This phenomenon has been coined the name Moore's law~\cite{schaller1997moore}. However, transistor miniaturization has hit a plateau in recent years due to technological limitations~\cite{waldrop2016chips, theis2017end}. Furthermore, advances in deep learning have heightened the demands for computation, memory, and energy. Addressing these issues requires new computing paradigms based on novel materials~\cite {nature_editorial2018big}.

Throughout history, new epochs of civilization have been shaped by the combined innovations of ideas and the methods used to express them. For instance, Johannes Gutenberg's invention of the printing press in 1436 enabled the rapid dissemination of knowledge through books, sparking a transformative global revolution. As computing methods and materials reach saturation, an urgent necessity arises for yet another revolutionary breakthrough. This revolution must encompass advancements in both computation as well as the materials that perform it. 

The animal world, having undergone millions of years of evolutionary optimization, is a significant inspiration for many modern marvels, such as the airplane. The human brain, in particular, exhibits remarkable pattern recognition abilities while also being energy-efficient. In the late 1980s, Carver Mead introduced the concept of neuromorphic computing, a brain-inspired approach designed to mimic how the brain processes information for computational tasks~\cite{mead1990neuromorphic, mead2020we}. While conventional computing developed in tandem with silicon and other semiconductor technologies, adopting a new computational paradigm would require innovative material technologies that naturally fulfill its unique needs. Spintronics-based materials, which utilize the spin degree of freedom, constitute one such class of materials showing promise as candidates for implementing neuromorphic computing. This chapter aims to motivate the potential of neuromorphic spintronics in addressing numerous challenges that plague contemporary computing technologies.

We begin this chapter by presenting the promise that spintronics-based materials hold as a platform for neuromorphic computing. This will be followed by exploring various application domains for spintronics-based materials. First, we consider the noisy yet error-tolerant aspect of our brain as an inspiration, and in that light, we discuss computing based on fluctuations using spintronics-based materials. Next, we discuss the implementation of neural networks, which are directly inspired by how neurons and synapses in our brain are configured to perform pattern recognition tasks. The subsequent chapter deals with another neuro-inspired algorithm called reservoir computing, and here we emphasize the usefulness of spintronics systems as suitable \textit{reservoirs}. In the penultimate chapter, we compare and contrast spintronics-based memories with other unconventional memory technologies. Lastly, we explore the future trajectory of neuromorphic spintronics, anticipating advancements in computational capabilities driven by new discoveries in the physical properties of spintronics materials.

\subsection{The promise of spintronics}
The field of electronics is based on using and manipulating the electron's charge. Like the charge, spin is another intrinsic property of the electron. Spintronics, short for ``spin-based electronics", is a field in physics and engineering that focuses on this degree of freedom for developing technology~\cite{vzutic2004spintronics, hirohata2020review}. Spin is a quantum mechanical property of any fundamental particle. It is associated with an intrinsic angular momentum of the particle, which allows for manipulation through various means, such as electric current, electric field, magnetic field, and light, among others.

The conceptual roots of spintronics can be traced back to the early 20th century with the discovery of the electron spin by Samuel Goudsmit and George Uhlenbeck in 1925~\cite{pais1989george}. However, it was not until several decades later that the potential for practical applications began to emerge; a significant breakthrough came in 1988 with the independent discovery of the giant magnetoresistance (GMR) effect by physicists Albert Fert and Peter Grünberg~\cite{baibich1988giant, binasch1989enhanced}. This discovery revealed how small changes in applied magnetic fields could lead to significant changes in the electrical resistance of certain materials. In the 1990s, the application of GMR in read heads for hard disk drives revolutionized computer data storage, dramatically increasing storage capacity and efficiency and ultimately earning Fert and Grünberg the Nobel Prize in Physics in 2007. At the end of the 20th century, further advances led to the development of the tunneling magnetoresistance (TMR) effect~\cite{moodera1995large, mathon2001theory, butler2001spin}. This led to the creation of Magnetoresistive Random-Access Memory (MRAM), a type of non-volatile memory that is based on Magnetic Tunnel Junctions (MTJs). The core physics behind TMR (in MTJs) and the GMR effect involves a three-layer magnetic junction, where a non-magnetic layer is sandwiched between two ferromagnetic layers. The non-magnetic layer provides an insulating tunnel effect and conduction pathway, respectively, for the MTJs and the GMR devices.
One of the ferromagnetic layers is a fixed reference layer, while the other is a free layer whose magnetic orientation can be altered. Data storage in MRAM is based on the principle of magnetoresistance – the change in electrical resistance depending on the magnetic alignment of these two layers. When the magnetic moments of the free and fixed layers are parallel, the device's resistance is low, representing a binary `1'. Conversely, the resistance is high when the magnetic moments are anti-parallel, representing a binary `0'. 

Another important milestone was the race-track memory~\cite{parkin2008magnetic}, where the motion of magnetic domain walls along nanowires is utilized. Here, the data is stored as magnetic bits along nanowire \textit{tracks}. These magnetic bits can be moved along the track by applying current pulses, enabling the reading and writing of data. The race-track memory technology, which is yet to be commercially available, aims to achieve a high storage density by packing magnetic bits close together on tracks, which can also be arranged three-dimensionally. In addition, racetrack memories provide the potential for faster and more energy-efficient memory systems compared to traditional technologies like Dynamic Random Access Memory (DRAM) and Flash. In the last decade, other magnetic textures, such as skyrmions, have been explored as magnetic bits~\cite{fert2013skyrmions, fert2017magnetic, tokura2020magnetic, everschor-Sitte2018perspective}, while antiferromagnets have demonstrated the potential for even greater transmission speeds~\cite{parkin2015memory}.

Unlike traditional electronics, which rely solely on the charge of electrons, spintronics also harnesses the spin of electrons to encode and process information. One key benefit is reduced energy consumption, as manipulating spin typically requires less energy than moving or storing charges. Additionally, spintronics devices can achieve faster processing speeds due to the intrinsically fast dynamics of spin states, enabling quicker data transmission and processing. This technology also holds potential for greater scalability, particularly in the context of miniaturizing devices, and offers enhanced stability and reliability, as spin currents or magnetized states are less susceptible to external perturbations such as electromagnetic interference. Furthermore, it provides many spin-dependent physical effects like TMR, spin-transfer torque (STT), spin-orbit torque (SOT), etc., which offer new opportunities for developing the next generation of storage and computing devices~\cite{ralph2008spin, ramaswamy2018recent}.

\subsection{A case for neuromorphic spintronics}
\label{sec:neurospin}
Conventional computers are built with silicon-based architectures and hardware, with algorithms tailored to the technological requirements and available resources. Contrary to that, neuromorphic computing is an approach that mimics the neural structure and functioning of the human brain to enhance computational efficiency and processing power. The inspiration could manifest at various levels: the architecture, the computing hardware (including the substrate), or even the software strategies and algorithms.

\begin{figure}
    \centering
\includegraphics[width=1.0\textwidth]{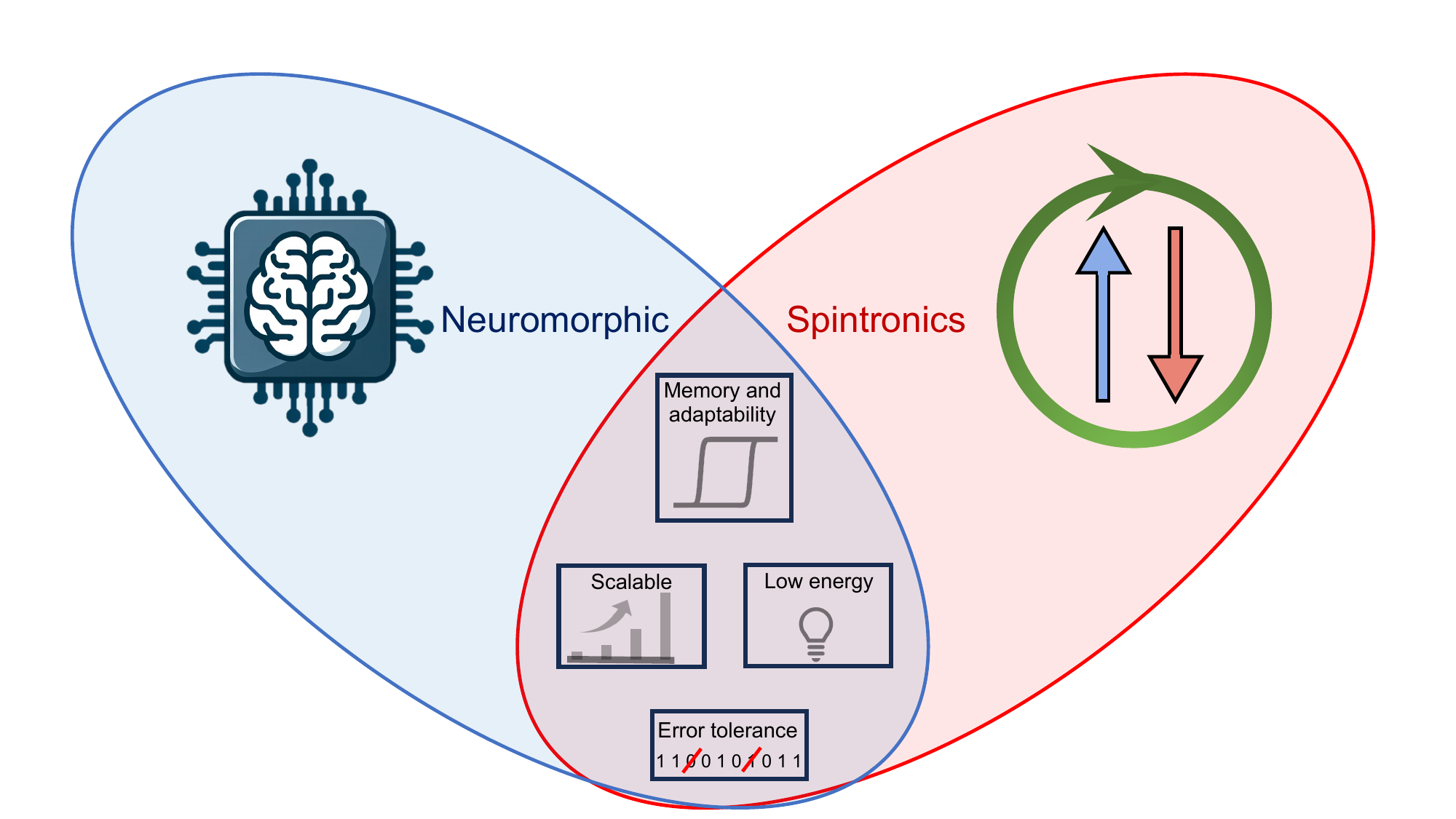}
    \caption{Neuromorphic computing with spintronics-based devices offers many characteristics such as memory and adaptability, scalability, low-energy computation, and error tolerance.}
    \label{fig:overview}
\end{figure}

Spintronics materials hold significant promise for neuromorphic computing due to their inherent characteristics, which are ideally suited for this novel computational paradigm. These characteristics can be broadly classified into four categories, as shown in Fig.~\ref{fig:overview}.

\begin{enumerate}
    \item \textbf{Memory and adaptability:} In the context of computing, memory refers to the ability to store values. In spin-based systems, a device can retain the magnetic state to which it is programmed, thus providing memory. An example is MRAM, where magnetic states are represented by parallel or anti-parallel alignments, thereby serving as memory. Compared to traditional charge-based memory technologies such as \changes{static random-access memory} (SRAM) and DRAM, spintronics-based memories are non-volatile, meaning they retain stored values even when powered off~\cite{upadhyay2019emerging, burr2017neuromorphic}. Non-volatility reduces energy consumption, as memory units do not require constant access in many practical applications. Furthermore, a single device can implement both memory and nonlinear behavior, unlike other memory systems that often require multiple transistors to emulate these features~\cite{grollier2020neuromorphic}.  
    
    In the human brain, both the abilities to memorize and to update memory are crucial for its functionality, highlighting the importance of adaptability. The brain can readjust its internal state by making its synaptic connections stronger or weaker and learning new information or modifying old memories. For spin-based systems, the magnetic state can be reversibly and controllably changed by expending a small amount of energy. The MRAM example uses a current (STT or SOT) to change the magnetic state from parallel to anti-parallel (or vice-versa). As we shall see in section~\ref{sec:3}, this ability to adapt, also referred to as \textit{plasticity}, is important for implementing neural networks with spintronics-based devices~\cite{kolb1998brain}.
    
    \item \textbf{Low energy:}
    To perform tasks such as pattern recognition, conventional computers consume orders of magnitude more power than the human brain, which uses around 20~Watts~\cite{attwell2001energy}. Our brain's efficiency stems from several factors, with its architecture being one of the primary reasons. Neurons (the basic computing/logic element) and the synapse (the basic memory element) are located physically together in the brain. In contrast, in conventional computers, the logic and memory units are physically separated, and a lot of energy is spent in shuttling information back and forth~\cite{horowitz20141}. This bottleneck of energy consumption in conventional computing paradigms, termed the \textit{von Neumann bottleneck}, can be alleviated using spintronics-based memory elements such as MRAMs~\cite{indiveri2015memory, burr2017neuromorphic}. Such novel memory technologies can be integrated with transistors that perform logic and arithmetic operations, thus facilitating in-memory or near-memory computing.

    Another factor contributing to low energy consumption in the human brain is its method of representing information. Neurons propagate information in the form of electrical voltage spikes. The presence of spikes, rather than a continuous signal, makes the representation of information sparse. This sparsity of activation can also be emulated using spintronics-based devices, which can implement leaky-integrate and fire neurons. Such examples will be described in more detail in section~\ref{sec:3}.

    \item \textbf{Scalability:} The human brain has a volume of just about 
    1,200~cm$^\textrm{3}$ but contains 10$^\textrm{11}$ neurons and 10$^\textrm{14}$ synaptic connections~\cite{cosgrove2007evolving, herculano2009human, pakkenberg2003aging}. The computational efficiency of our brain is partially attributed to this massive connectivity. Although achieving such high connectivity in hardware remains challenging due to limitations in 2D and 3D device fabrication techniques, spintronics devices, with their miniaturization capabilities, can enable significant advances in device integration. The underlying mechanisms of spintronics devices maintain functionality even when device dimensions are reduced to a few nanometers~\cite{finocchio2021promise}. In contrast, charge-based effects in traditional electronics become unreliable at such small scales due to quantum tunneling and electron leakage~\cite{bhatti2017spintronics}.

    The huge connectivity in the brain is also responsible for its massively parallel computation. The ability to process multiple information streams in parallel is critical for the brain's adaptability and functionality, underpinning everything from basic survival instincts to higher cognitive functions. The scalability offered by neuromorphic spintronics would allow such parallelization to be possible in hardware as well, where tasks can be divided across a vast array of processing units, enabling the system to handle complex, dynamic processes efficiently. 
    
    Realistically, replicating the extensive connectivity and dynamic reorganization of our brain in hardware is not feasible. Instead, adopting a modular approach allows specific hardware components to fulfill distinct roles regulated through software. This strategy effectively addresses the challenges of connectivity and the dynamic reconfiguration of synaptic connections.
 
    Spintronics devices easily integrate with existing CMOS technology, aligning well with ongoing advancements in silicon-based systems~\cite{finocchio2021promise}. This integration combines the best features of both spintronics and silicon-based technologies, advancing overall technological development.
    
    \item \textbf{Error tolerance:} While the brain is capable of pattern recognition at a low energy budget, it is also quite noisy~\cite{faisal2008noise, rolls2010noisy}. The sources behind this noise can range from fluctuations in ion channel activities and synaptic transmission variabilities to spontaneous neuronal firings. Ion channels, for instance, might open and close randomly, leading to variability in the membrane potential of neurons~\cite{faisal2005ion, faisal2008noise}.

    Globally, the brain represents a highly complex, nonlinear network where small perturbations can result in large-scale changes. Despite these fluctuations, the brain excels at robust pattern recognition. This capability largely stems from the substantial redundancy within the brain; multiple neurons and pathways often represent the same or similar information, allowing other pathways to compensate when one is affected by noise. The vast scalability potential of spintronic devices enables them to mimic this redundancy, enhancing their capacity for reliable information processing and pattern recognition.

    Furthermore, charge-based memories, particularly those with miniaturized components, are more sensitive to environmental factors such as radiation exposure, leading to higher error rates. On the contrary, spintronics-based devices are often more robust to radiation, which can have significant advantages for extra-terrestrial applications~\cite{ren2012radiation}.
\end{enumerate}

While the complete understanding of brain function remains elusive, it is suggested that the inherent stochasticity within the brain could play a crucial role in enhancing its computational efficiency~\cite{mori2002noise, deco2009stochastic}. Considering this, we delve into discussions around computing with fluctuations and its spintronics implementations.

\section{Computing based on fluctuations}
\label{sec:2}

Usually, in the realm of computation and other practical applications, stochasticity and fluctuations are a major nuisance, necessitating extensive algorithmic and instrumental machinery to minimize their impact. The sources of noise are often intrinsic to the system, either due to the inhomogeneities present or thermal fluctuations. This inherent noise can be harnessed to perform computations in various ways, which will be presented in this section. Each of these computational paradigms has specialized use cases based on their properties.

\begin{figure}
    \centering
\includegraphics[width=0.9\textwidth]{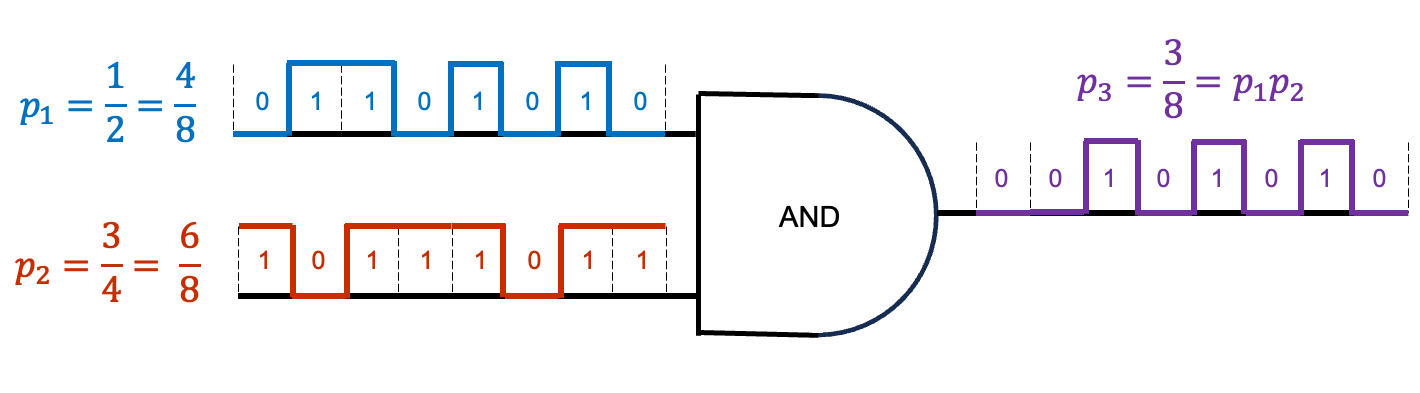}
    \caption{Illustration of the operating principle of stochastic computing using the example of multiplication of two numbers $p_1$ and $p_2$.
    The (in general approximate) result of the multiplication $p_3$ is obtained by connecting uncorrelated random bitstream representations of $p_1$ and $p_2$ using an AND gate.}
    \label{fig:stochastic_computing}
\end{figure}

\subsection{Stochastic computing}
Originally introduced in the 1960s, stochastic computing emerged as a cost-effective computing scheme that is perfectly suited for computations where the accuracy could be traded in favor of the speed~\cite{gaines1969stochastic, alaghi2013survey}. A common example in everyday life is a vacuum cleaning robot designed to quickly clean the floor, where millimeter-level precision is not a concern. The core concept of this method is to compute with probability values, i.e., numbers between 0 and 1, which can be approximated by random uncorrelated streams of bits, where the proportion of ones approximates the numerical value. Computational operations are performed directly on the bitstreams. For instance, multiplying two numbers is achieved by feeding their respective bitstreams into an AND gate, as exemplified in  Fig.~\ref{fig:stochastic_computing}. Other arithmetic calculations can also be done using gates and multiplexers~\cite{alaghi2013survey}.

The primary advantage of this method is that it circumvents the need for the complex and energy-intensive circuits typically used for binary arithmetic calculations. Instead, simple gates are 
employed, offering a solution with substantially lower energy consumption and reduced space requirements. Furthermore, this approach is more error-tolerant since single erroneous bit flips in the bitstreams only mildly affect the results compared to ordinary bit-based calculations.

Despite its advantages, such as inherent error tolerance, simplicity of design, and energy efficiency, this computing paradigm suffers 
from the trade-off between accuracy and calculation time. This is because the mathematical basis of stochastic computing is the law of large numbers, which states that the sample mean of a random variable converges to the true mean in the limit of a large number of samples~\cite{hsu1947complete}. Thus, the law of large numbers guarantees that the longer the bitstreams, the more accurate the representation of the desired probability value is on average~\cite{alaghi2017promise}.

Moreover, a key requirement for computing with bitstreams is that they need to be uncorrelated. The easiest way to understand this is by using the example of multiplication of two of the same numbers. For two identical input bitstreams representing this number (i.e.\, when they are maximally correlated), the output of the AND gate will be exactly the same as the input streams. Thus, the calculation result will not be the product of the input's probability values but simply the probability itself.

Spintronic devices have been proposed and demonstrated as effective generators of uncorrelated bitstreams, functioning as random number generators. For example, MTJs with a low energy barrier between the parallel and anti-parallel states, also called superparamagnetic tunnel junctions, have been proposed as random number generators~\cite{vodenicarevic2017low, fu2021overview}. Here, thermal fluctuations cause random switchings between the two magnetic states, which can be interpreted as the bits 0 and 1.

Similarly, magnetic textures such as skyrmions reshufflers can be leveraged for generating uncorrelated bitstreams~\cite{pinna2018skyrmion, zazvorka2019thermal}. This is achieved by representing the bitstreams as skyrmions moving under a constant current into two separate chambers. The value of each bit (0 or 1) determines which chamber receives a skyrmion. Once inside a chamber, the skyrmions undergo random thermal diffusion, effectively causing them to \textit{reshuffle}. The output is read similarly, where receiving a skyrmion from the first chamber indicates a 0, and from the second chamber indicates a 1. This method ensures that while the probability value encoded in the bitstreams remains unchanged, the correlation among the bitstreams is decoupled.

\begin{figure}
    \centering
    \includegraphics[width=1.0\textwidth]{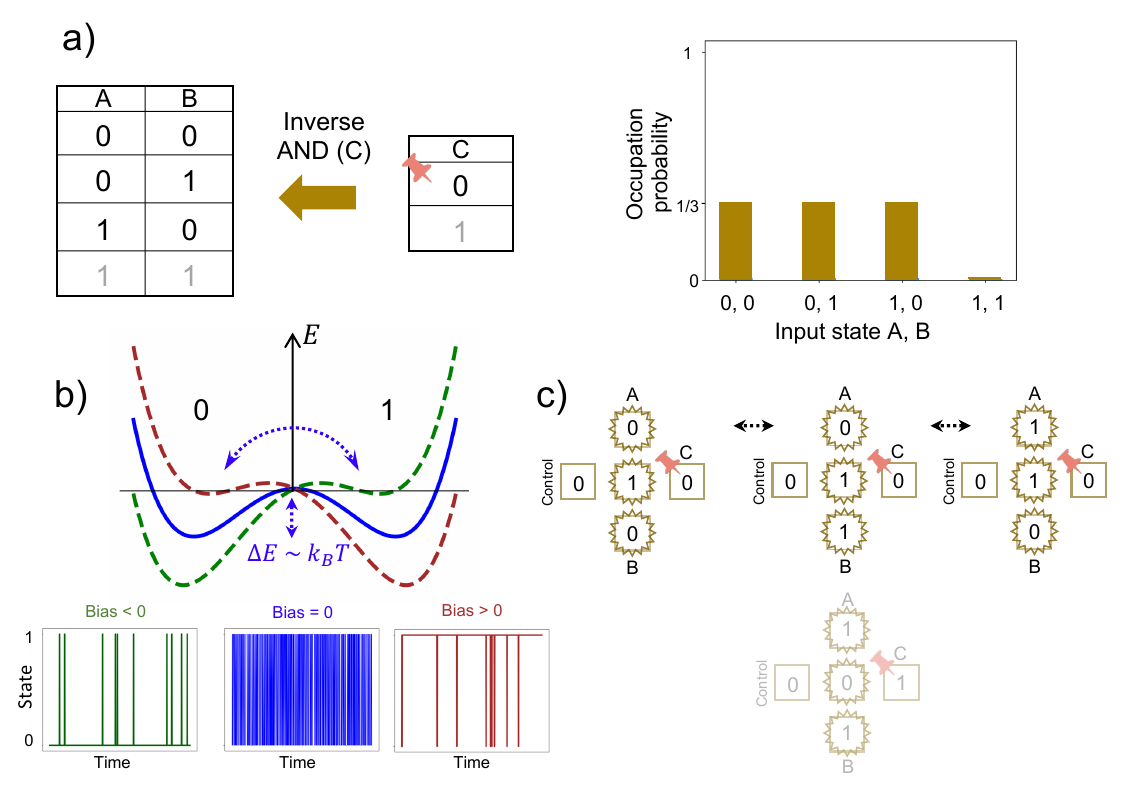}
    \caption{Illustration of inverse computing exploiting p-bits. (a) Schematic of the general idea behind inverse computing showing an example of an AND gate. The inverse for the output ``0'' is not unique. Three different inputs (00, 01, 10) correspond to the output ``0''. An ideal inverse computing scheme utilizing fluctuations yields all three input possibilities with equal probability. (b) Key characteristics of a p-bit: the bistable system has two energy ($E$) minima at states representing 0 and 1 separated by an energy barrier ($\Delta E$), which is of the order of $k_B T$, where $T$ is the ambient temperature. Under zero bias (blue), the p-bit fluctuates equally between the two states. Under positive/negative bias (brown/green), the energy landscape is tilted, and the 1/0 state is favored, thus changing the represented probability value. \changes{(c) Operation of a p-bit with nanomagnets: a majority gate setup can be used to implement an AND gate using p-bits. When the output bit C is fixed (indicated by the light red pin) to the 0 state, the A and B bits fluctuate between the three possible states. The jagged lines for the bits indicate that they can toggle between the two states, 0 and 1.} }
    \label{fig:invertible_computing}
\end{figure}

\subsection{Inverse computing}
Typically, computing involves producing the correct output from a given input through systematic processing. Identifying the inverse relationship — tracing back from the output to the original input — can also be particularly useful in applications such as integer factorization and invertible logic~\cite{borders2019integer}. Inverse computing is inherently challenging as mathematically, often no unique inverse exists, i.e.\,  multiple inputs often correspond to a single output. For instance, as illustrated in Fig.~\ref{fig:invertible_computing} (a), output 0 of the AND gate corresponds to three different input states. 

Thus, the entire input space must be searched to find all inputs that produce the desired output, making this process typically far more resource-intensive than standard forward computing. Gates that can function in reverse would be ideal; however, an additional challenge is that the input space often also grows exponentially with the number of inputs.

\changes{One approach is to use fluctuations to explore the entire solution space in parallel, increasing the likelihood of finding all possible input states with equal probability. By arranging fluctuating elements to favor the correct solution, the system becomes statistically more likely to settle into the desired state. This concept is illustrated through a specific example below.}

\changes{A promising building block for the basic fluctuating element is the p-bit.} 
A p-bit or a probabilistic bit fluctuates between 0 and 1~\cite{camsari2019p, camsari2017stochastic} in contrast to a classical bit, which takes a definite value, i.e.\, either 0 or 1. The p-bit represents the probability value as the fraction of time it spends in the state labeled as 1 relative to the entire operational period. In principle, any bistable physical system where the switching probabilities can be tuned by an external bias can be used to implement p-bits, as shown in Fig.~\ref{fig:invertible_computing} (b). Here, the energy needed to switch between the two minima must be in the order of the thermal energy, such that the p-bit naturally fluctuates between the two states at ambient temperature. \changes{In the figure, the 0 and 1 p-bit states are represented by two distinct physical states of the system. For a spintronics-based system, the magnetization order parameter, for example, can represent the state of the system.} Ideally, under unbiased conditions, the p-bit stays in each of the two states with an equal probability and frequently switches between them. Applying an external bias alters the energy landscape, making the system more likely to reside in one of the two states. A possible spintronic implementation of p-bits utilizes superparamagnetic tunnel junctions. Here, the low energy barrier between the parallel and anti-parallel alignment of the magnetic layers can be overcome by thermal fluctuations, allowing to switch between the two states~\cite{camsari2017implementing}. The biasing of the states can be achieved through various methods, such as an external magnetic field, bias voltage, STT currents, or SOT currents~\cite{chowdhury2023full, yin2022scalable}. 

\changes{In Fig.\ref{fig:invertible_computing} (c), we present a specific example of how the p-bit-based inverted AND gate can be implemented. This type of logic utilizes majority gates with nanomagnetic islands that are anti-ferromagnetically coupled\cite{imre2006majority, eichwald2014majority}. The control bit governs the logic operation and is set to ``0" for the AND function. For the inverse AND gate, the output bit C is pinned (represented by a red pin in Fig.\ref{fig:invertible_computing} (c)), while the three p-bits in the middle column are free to fluctuate (shown by zigzag lines). When the output C is fixed at state ``0," the central island energetically favors the ``1" state due to antiferromagnetic coupling between neighboring islands. If the next-nearest neighbor coupling is also antiferromagnetic, the states of p-bits A and B become non-unique and fluctuate between the three different input states that correspond to the AND gate’s ``0" output (first row of Fig.\ref{fig:invertible_computing} (c)).}

\changes
{This simple example also highlights several challenges in using a majority gate-based approach for inverse computing. First, the energy landscape and interactions must be precisely designed to ensure accurate gate functionality. Second, achieving balanced inverted gate operations—where all correct input states have equal probability—will be quite difficult. In addition to refining the interactions and structure geometry, individual p-bits can also be biased to create more balanced gates.}

In summary, the primary technological advantages of inverse computing with p-bits are the parallel exploration of the input space and ultra-low energy consumption, driven by the inherent stochasticity introduced by thermal fluctuations. Another possibility for efficiently solving inverse problems is proposed in the context of memcomputing. In this alternative computing paradigm, memory elements are used directly for processing by self-organizing logic gates~\cite{di2018perspective, traversa2017polynomial}. There have also been proposals to implement memcomputing using spintronics-based devices, enabling the execution of inverse computations~\cite{finocchio2021promise, finocchio2023roadmap}.

\begin{figure}
    \centering
    \includegraphics[width=0.9\textwidth]{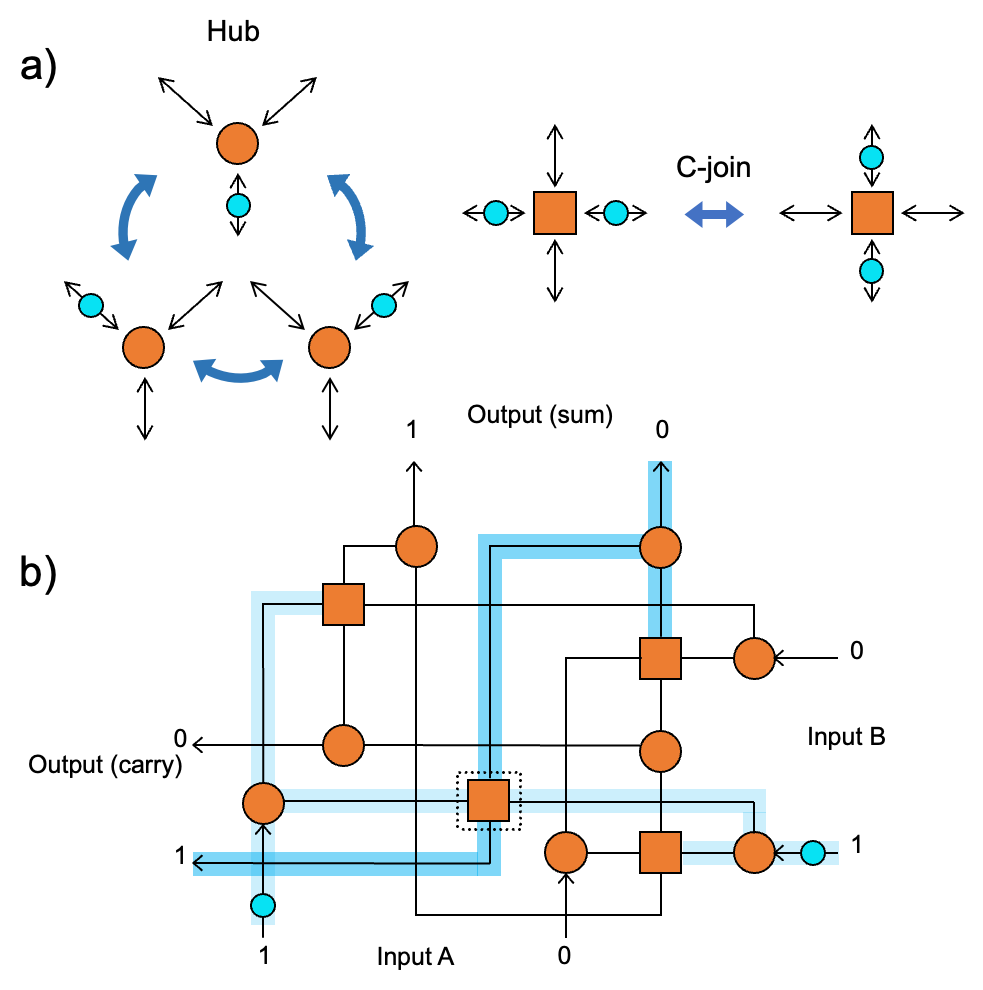}
    \caption{Illustration of the basic principles of token-based Brownian computing. The tokens (cyan-colored circles) perform a random motion along the wires (black) and are influenced by the circuit elements (orange) according to their functions.
    a) One basic set of primitive elements consists of a Hub (circle) and a C-join (Square). The Hub is a triple junction that allows the token to move freely along any of the three connecting wires. The C-join is a four-way intersection that requires two tokens to enter the element from different directions, and it releases them along the other two directions.
    b) Implementation of a half-adder circuit using hubs and C-joins. In the specific example, the calculation of the addition of two bits (Input A and B) with the value 1 is shown. The wires marked in (light) blue indicate along which wires the tokens are allowed to move (before) after passing the C-join. Adding two bits with the value 1 thus leads to the desired result of carry 1 and sum 0, which is where the tokens exit the circuit and can be detected. The C-join that mainly contributes to this operation is marked with a dotted square. }
    \label{fig:token_based_Brownian_computing}
\end{figure}

\subsection{Token-based Brownian computing}

In magnetic p-bits, the stochasticity originates from the fluctuations in the magnetic states. The thermal Brownian motion of nanomagnetic solitons can also be harnessed for computation in a paradigm called token-based Brownian computing~\cite{nozaki2019brownian}. 
Whereas for stochastic computing and p-bits, the stochasticity resulted in generating bits, in token-based Brownian computing, the thermal fluctuations are utilized to randomly propagate tokens, which can physically explore the search space of a task. In this context, a token refers to a solitary discrete object whose presence or absence is interpreted as the signal. The computational task or logical operation is mapped into a circuit where the tokens undergo Brownian motion along certain paths guided by a few basic, resource-friendly elements~\cite{peper2009exploiting, lee2010efficient}.

A minimal example set of basic elements for performing logical operations consists of a Hub and a C-join~\cite{peper2009exploiting}, as shown in Fig.~\ref{fig:token_based_Brownian_computing}(a). The Hub element is a trijunction, i.e.,\ it consists of three bidirectional wires enabling the token to move randomly along any of the connected wires. The C-join is a four-way intersection that acts as a signal synchronizer: when two tokens enter the element from different directions, the C-join releases them along the other two directions.

Fig.~\ref{fig:token_based_Brownian_computing} (b) shows a possible realization of a half-adder circuit exploiting only the Hub and the C-join. This basic example illustrates when token-based Brownian computations are useful; the calculation relies only on the thermal motion of the tokens, and almost no external power is needed (besides the ultra-low power required to activate the C-joins.) The disadvantage, however, is the potentially very long time required to obtain the calculation result, as there is no guaranteed time until the tokens reach the output lines. Adding ratchets, another type of basic element, at suitable locations in the circuit can accelerate the convergence to the solution. Ratchets limit the token's movement along a certain direction, i.e., it cannot go backward after a token has passed a ratchet. In the half-adder shown in Fig.~\ref{fig:token_based_Brownian_computing} (b), they can, for example, be placed after the C-joins.

Using spintronics systems, an energy-efficient implementation of the tokens and circuit primitives can be achieved. Skyrmions, or other topologically protected textures that undergo such Brownian motion, can play the role of the tokens~\cite{nozaki2019brownian}. The Hub has been demonstrated using circuit geometry in magnetic materials~\cite{jibiki2020skyrmion}, the C-join through voltage-controlled magnetic anisotropy (VCMA) effects~\cite{nozaki2019brownian}, and the ratchet by employing either a VCMA gradient or specially designed wire geometries~\cite{wang2018efficient, souza2021skyrmion}. An additional advantage of spintronics systems is that they allow to control and enhance the diffusion of magnetic solitons and, in particular, skyrmions via different mechanisms~\cite{goto2021stochastic}.

\section{Spintronics-based neural networks}
\label{sec:3}

Human intelligence relies not only on the capacity to perform computations but also on the ability to recognize patterns. Artificial neural networks (ANNs), inspired by the neural connections in the brain, excel at pattern recognition, driving the majority of the Artificial Intelligence (AI) advancements in the present era~\cite{lecun2015deep, goodfellow2016deep}. 
Currently, ANNs are used across various fields, including image recognition, natural language processing, the study of protein folding~\cite{guo2016deep, deng2018deep, jumper2021highly} and its applications have even pervaded many aspects of our daily lives~\cite{alzubaidi2021review} (see also chapter \textcolor{red}{(chapter by Tobias Wand)}).

An ANN essentially serves as a function approximator with a huge number of tunable parameters (e.g.\ GPT-3 has 175 billion parameters~\cite{zhang2021commentary}), which are tweaked to learn a good mapping between an input and its target label~\cite{cybenko1989approximation}. 
The rapid increase in network size over the recent years and the development of more complex architectures have significantly increased the energy requirements for training and running these models on conventional hardware. 
This energy consumption has adverse environmental consequences, prompting the search for resource-friendly substrates for implementing neural networks~\cite{patterson2021carbon}. 

In this chapter, we address the hardware implementation of neural networks with spintronics.
 Before taking a deeper dive into the implementation of ANNs using spintronics and the advantages that it delivers over conventional hardware, we summarize its basic functionalities.

\subsection{Basic mode of operation of ANNs}
ANNs consist of two main components: the neurons (or neuronal activations) and the synapses (or synaptic weights) interconnecting the neurons.
A deep neural network features multiple neuron layers that transform input into output via hidden layers.
Specifically, in a supervised learning setting, an input for a layer is fed into the neurons ($x_i$). Then, using the synaptic weights ($w_{ij}$) attached to the neurons, 
the $j^{\textrm{th}}$ neuronal activation $a_j$ in the next layer is computed as $a_j = f\big(\sum_i w_{ij}x_i$\big). Here, the activation function $f$ introduces non-linearity into the network, which is indispensable for solving complex problems.

For the physical realization of ANNs, generally, factors such as energy efficiency, speed, error tolerance, compatibility with CMOS technology, endurance, low area overhead, and fabrication cost are crucial considerations. In addition, there are specific requirements for the synaptic and neuronal elements, as explained below.

\subsubsection{Requirements for Synapses}
The synaptic weights serve as the parameters that characterize the neural network model~\cite{lecun2015deep}.
They can be regarded as memory elements whose main task is to store the value of the weight in a robust, non-volatile manner for inference. To enable training with materials, the physical properties of the memory elements—typically resistance or conductance—that correspond to weights must be precisely adjusted according to updates from the training algorithm.
Resistance-based memories (also called memristors) offer a natural way to implement synapses in the following way~\cite{strukov2008missing}: if a weight $w_{ij}$ is encoded as the conductance $g_{ij}$ of a device at the junction between the $i^{\textrm{th}}$ input and $j^{\textrm{th}}$ output, and if the input and output are represented respectively by voltage $V_i$ and current $I_j$, then using the Ohm's law and Kirchoff's current law we can write the output current as $I_j = \sum_i g_{ij}V_i$.

Thus, the conductance of any physical system could be used to implement the basic functionality of synapses. However, for neural networks, additional properties are required that current hardware does not provide. Firstly, it is essential that synapses are non-volatile, meaning the memory elements retain their programmed state even when the network is not powered. Secondly, overcoming the \textit{von Neumann bottleneck} necessitates the co-location of computing and memory elements. It is important to note that, in a neural network, synapses outnumber neurons, making them the primary bottleneck for in-memory implementation and, thus, a critical focus for hardware development.

\subsubsection{Requirements for Neurons}

The neurons are the basic nodes or computational elements of the network. Their main purpose is to implement the accumulation of the inputs from the previous layer multiplied by the connecting weights and the nonlinear activation function $f$. Such computing neurons, which typically take continuous real values, can be implemented using a few transistors. However, the implementation with transistors usually has a high area overhead, is energy-inefficient, and suffers from device-to-device variability.

As opposed to the neurons of the simple feed-forward ANNs, the neurons in our brain are dynamic, excitable cells. They accumulate the incoming electrical signal from other neurons in the form of spikes in a leaky manner. While integrating the signal, when the membrane electric potential reaches a certain threshold, it releases an electrical voltage spike. Following this discharge, the membrane potential returns to its baseline state and enters a refractory phase. Based on this idea, Spiking Neural Networks (SNNs) have been proposed as another more bio-realistic variant of neural networks where the neurons communicate with each other via sparse, discrete spiking signals~\cite{tavanaei2019deep}. Although more difficult to train in terms of speed and accuracy, SNNs are more energy-efficient because of the sparse nature of their representation~\cite{blouw2019benchmarking}. To emulate SNNs, the neurons must possess a mechanism for generating spikes. Typically, the mechanism used for producing the spikes is called the leaky integrate and fire neurons. They integrate the input signal and only fire when this integrated value reaches a certain threshold~\cite{liu2001spike}.

\subsection{Spintronics-based building blocks for neural networks}
Spintronics-based hardware is inherently more energy-efficient than current transistor technology, and research has shown that spintronics technology is congruent with implementing the different building blocks of neural networks. In the subsequent text, we present examples of both spintronics-based synapses and neurons.

\subsubsection{Spintronics-based synapses}
Spintronics-based synapses exploit the magnetic state of a physical system to store the synaptic weight. Besides the general advantages of spintronics systems mentioned in section~\ref{sec:neurospin}, spintronics systems offer the reliable and efficient reading and writing of non-volatile magnetic states using effects such as anisotropic magnetoresistance, STT, and SOT. This enables a variety of concepts for spintronics-based synapses, including the following.

For example, spintronics synapses using MTJs have been proposed, and the junctions can be utilized in various ways. An instance of this is using them as binary synaptic elements~\cite{jung2022crossbar}, where the two (parallel and anti-parallel) magnetic states represent the two binary states. In addition, the stochastic switching property has been leveraged as stochastic synapses~\cite{vincent2015spin} and even as synapses controlled by radio frequency signals~\cite{leroux2021hardware}. Another class of spintronic synapses is based on magnetic textures like magnetic skyrmions or domain walls.
Here, for example, the number of skyrmions in a region dictates the conductance and, consequently, the synaptic weight~\cite{huang2017magnetic, song2020skyrmion}. For magnetic domain walls, it has been proposed to use the position of the wall as synaptic weight, as this can also change the conductance~\cite{bhowmik2019chip, liu2021domain}. Furthermore, antiferromagnetic materials have been identified to have magnetization-switching properties that can be used for synaptic applications~\cite{miron2011perpendicular, fukami2016magnetization}. 

\subsubsection{Spintronics-based neurons}

Various spintronics-based implementations of neurons have been proposed to utilize magnetic states, as discussed in this subsection. In particular, they truly shine when implementing leaky integrate and fire neurons. 

Spin-torque nano-oscillators, a type of MTJ, exhibit spontaneous microwave oscillations when driven by direct current. Their memory-like oscillation amplitudes mimic neuronal leaky integration, while the nonlinear response of voltage oscillation amplitudes to input current or field enables the direct implementation of activation functions~\cite{kiselev2003microwave, torrejon2017neuromorphic}. Another example is the superparamagnetic tunnel junction, which is an MTJ with a very low energy barrier between the parallel and anti-parallel states. Here, the switching rate can be tuned by magnetic fields and spin torque effects, enabling them to be used as neurons in a scheme of computing called population coding~\cite{locatelli2014noise, mizrahi2018neural}. 

Neurons, like synapses, have also been proposed with magnetic textures. These magnetic textures, such as skyrmions and domain walls, offer the advantage of being controllable through energy-efficient methods, such as applying a low current. This capability allows them to function as carriers of information, akin to neurons, effectively. Furthermore, these textures, owing to their nanoscale dimensions, can be prone to noise either from pinning sites or from thermal fluctuations~\cite{hayward2015intrinsic}. This feature holds promise in emulating the functionality of stochastic neurons. Leaky integrate-and-fire neurons based on magnetic textures have also been proposed for skyrmionic or domain wall systems. In these systems, the gradual accumulation and cumulative motion of the solitons ultimately lead to switching of the magnetic state~\cite{sharad2012spin, hassan2018magnetic, chen2018magnetic}. This type of leaky neuron has also been proposed in synthetic antiferromagnetic systems~\cite{wang2023spintronic}.

Moving forward, we delve into reservoir computing, where we highlight the diverse functionalities of the spintronics-based approach

\section{Reservoir computing with spintronics}
\label{sec:4}

Reservoir computing exploits the inherent nonlinear complex dynamics of a system (called the reservoir) to simplify a classification or prediction task. For this, the reservoir project inputs into a higher dimensional space where the classification or prediction task reduces to an easy (e.g.\ linear) separation task. The first reservoirs were implemented by recurrent neural networks with random but fixed synaptic weights, such that only the last output layer was trained~\cite{lukovsevivcius2009reservoir}.
In recent years, it has become apparent that many physical systems are well-suited as reservoirs. The emerging field of physical reservoir computing leverages the properties of physical systems to enhance computational performance~\cite{nakajima2020physical}.

Reservoir computing has been discussed in detail in chapter \textcolor{red}{(chapter by Michael te Vrugt)}, but here we want to highlight some key features that a physical system has to have to function as a reservoir computer~\cite{everschor2024topological, lee2023perspective}.

\begin{itemize}
    \item \textbf{Non-linearity:} This refers to the input undergoing a nonlinear transformation due to the reservoir.
    \item \textbf{Complexity:} The term complexity relates to the reservoir's ability to effectively project inputs into a space of higher dimensionality, where ``high dimensionality'' refers to a reservoir possessing considerably more degrees of freedom than the inputs have.
    \item \textbf{Short-term or fading memory:} The reservoir's ability to process the input signal's temporal history is characterized by its short-term memory attribute. This attribute prioritizes recent inputs while retaining information from earlier ones, with the output being influenced by past inputs, though their impact gradually fades over time. This fading memory is vital for the system to be more responsive to recent inputs, enabling it to adapt to new input patterns or discard outdated information.
    \item \textbf{Reproducibility:} The reservoir should yield identical responses to identical inputs, provided it has been reset between each input. This attribute, although trivial for software implementations, is a crucial prerequisite for physical reservoirs.
\end{itemize}
The strength of physical reservoir computing lies in its flexibility, allowing a variety of systems that meet the aforementioned criteria to serve as the reservoir. 

\subsection{Magnetic textures for reservoir computing}

Magnetic systems ideally fulfill all criteria to serve as suitable reservoirs due to their inherently nonlinear and complex responses to external stimuli. Also, their ability to ``forget" information over time -- a characteristic known as fading memory -- arises from Gilbert damping and other mechanisms that dissipate energy in the magnetization dynamics. Additionally, their compatibility with CMOS technology means magnetic reservoirs can seamlessly integrate into modern electronic devices.

\begin{figure}
    \centering
    \includegraphics[width=1.0\textwidth]{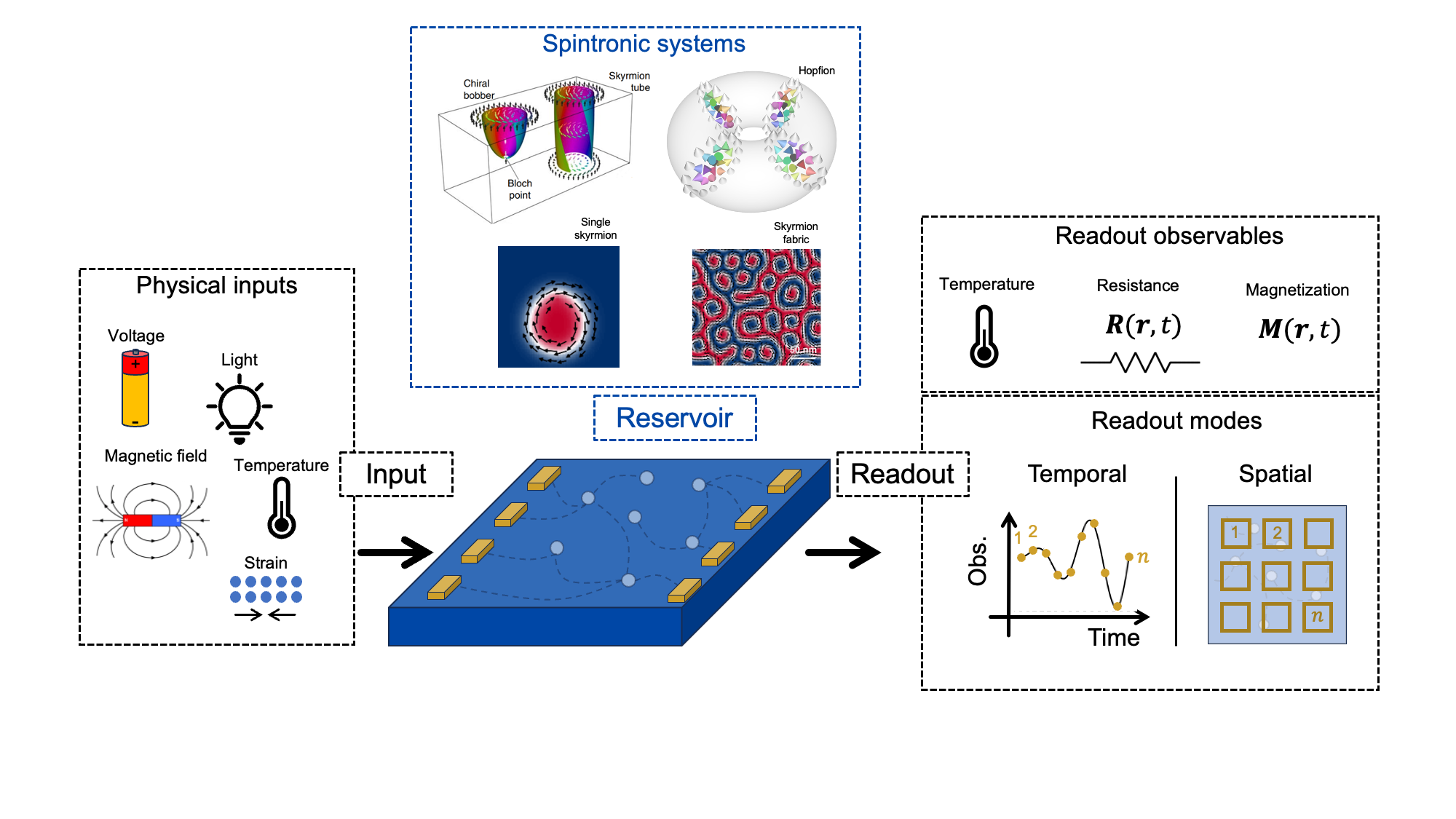}
    \caption{Illustration depicting various physical operational and readout modes of reservoir computing. The input can be supplied to the reservoir via different physical quantities like voltage, light, magnetic field, temperature, and strain. A plethora of different spintronics systems can play the role of the reservoir. The reservoir states can be read by measuring different physical quantities like temperature, resistance, or magnetization. Furthermore, the mode of reading out the reservoir state could be spatial or temporal. (Figure adapted from figures in~\cite{everschor2024topological, prychynenko2018magnetic, bourianoff2018potential, zheng2018experimental}). Hopfion figure created by Ross Knapman.}
    \label{fig:RC}
\end{figure}

Physical reservoir computing has been demonstrated with common magnetic building blocks such as MTJs~\cite{torrejon2017neuromorphic} and spin-vortex nano oscillators~\cite{markovic2019reservoir}. However, despite their utility, these systems fall short in the complexity required for effective reservoir computing, especially in terms of scaling to more difficult tasks. Enhancing their complexity, for instance, through time multiplexing, is necessary to leverage them as reservoirs~\cite{rohm2018multiplexed, cucchi2022hands}. Conversely, magnetic systems with more complexity like dipole-coupled nanomagnets~\cite{nomura2019reservoir, gartside2022reconfigurable}, spin wave-based reservoirs~\cite{nomura2019reservoir}, magnetic metamaterials~\cite{vidamour2023reconfigurable}, skyrmion fabrics~\cite{prychynenko2018magnetic, bourianoff2018potential, pinna2018reservoir, sun2023experimental, leeMK2022reservoir, leeMK2023handwritten, msiska2023audio}, and other complex spin textures~\cite{bechler2023helitronics, lee2023btask, lee2023perspective} have been used as physical reservoir computers. Other topological spin textures, such as anti-skyrmions, hopfions, and dislocations~\cite{azhar2022screw, tang2021magnetic, wang2019current, kent2021creation, stepanova2021detection}, are gaining attention and in the future, can potentially serve as reservoir. Such diversity in terms of options for the reservoir is shown in Fig.~\ref{fig:RC}.

Furthermore, the spin-textures and magnetic states can be excited by a wide range of different physical inputs like voltage, light, temperature, magnetic field, and strain, any of which could be used as the input for the reservoir~\cite{vedmedenko2020the, lee2023perspective, grollier2020neuromorphic}. This flexibility in inputs offers a critical advantage for magnetic reservoir computers, as it allows optimization for speed, accuracy, or power efficiency based on the specific use case, leveraging the unique benefits of each input mode. Furthermore, if a sensor outputs these physical properties, it can be directly integrated with the reservoir, eliminating the need for conversion to an intermediate quantity.

After the inputs excite the reservoir, it has to be read out for the final trained layer. This, too, can be done via different physical observables: the input drive can locally change the temperature, magnetization, and, in turn, the electrical resistance, which can be measured~\cite{everschor2024topological, tanaka2019recent}. 
In addition to various physical observables, the readout can be done in multiple ways. For instance, the reservoir's response can be monitored over time~\cite{pinna2018reservoir}. Alternatively, or in addition, spatial sampling of the reservoir can be employed for readouts~\cite{msiska2023audio}.

The readout values are then utilized to train the weights of the final linear layer, which produces the output predictions. Currently, the training of these weights occurs on an external device in physical reservoir computing schemes, and further research is needed to explore how this process can be implemented directly within the material itself. Doing so could significantly reduce data transfer and the need for additional hardware, leading to faster computations, miniaturization, and reduced energy consumption. Additionally, more research needs to be done to establish a fair and reliable method for comparing the performance of different reservoirs. This includes distinguishing the impact of pre-processing and understanding the effects of artificially increasing a system's complexity.

\section{Memory technologies: spintronic implementations and beyond}
\label{sec:5}

\changes{The current memory technology landscape is primarily dominated by SRAM, DRAM, and NAND flash. SRAM utilizes bistable flip-flop circuits made of transistors to store data as long as power is provided, making it fast and ideal for cache memory. DRAM, on the other hand, stores data in capacitors and transistors, requiring periodic refreshing to retain information, and is commonly used as main memory in computers due to its higher density and lower cost. NAND flash memory, composed of floating-gate transistors, stores data by adjusting charge levels in memory cells, allowing it to retain data without power. It is widely used in SSDs, USB drives, and memory cards because of its high storage capacity and durability. All of these memory technologies depend on charge storage to represent a memory state. However, a new class of memory technologies is emerging that relies on resistance instead of charge, including spintronics-based memories, which stand out for their non-volatile nature, fast operation, and scalability, among other benefits.} The long-term vision for this class of memories is not to fully replace existing silicon-based technology but to serve specific use cases, such as embedded applications for edge devices. In such applications, energy efficiency, speed, error tolerance, integrability with existing technology, and low area overhead are considerations of paramount importance. In terms of implementing artificial intelligence algorithms like ANNs, the switching time and endurance are important parameters. For edge learning applications where the system has to be trained locally, fast and reliable switching is necessary as the network synaptic parameters are updated very frequently. In the context of memory technologies, ``endurance" refers to the durability or lifespan of a memory device, specifically measured by the number of write-erase cycles it can withstand before its performance degrades to an unacceptable level. For real tasks, the networks need to be trained for a very long time, and hence, the endurance of the devices needs to be high.

In addition to spintronic-based memories (such as MRAMs mentioned above), three other main types of memory technologies are attracting a lot of attention: oxide filamentary materials (OxRAMs), phase change memories (PCRAMs), and ferroelectric memories (FeRAMs)~\cite{ielmini2018memory}. All of these technologies exploit the non-volatile physical properties of a material to encode information.

OxRAMs change their resistance by the formation and dissolution of a conducting filament made with oxygen vacancies within an insulating material~\cite{dittmann2021nanoionic, beck2000reproducible}, whereas in PCRAM the material undergoes a phase transition that leads to differences in resistances~\cite{yamada1991rapid, wong2010phase, burr2010phase}. Some of these technologies have matured to become commercially available~\cite{zha2017recent, lee2018embedded}. In FeRAMs, a relatively new form of memory, the dielectric polarization of the material changes with the application of voltage, which in turn is measured as a change in resistance using a FeFET (Ferroelectric field effect transistor)~\cite{mikolajick2001feram, boscke2011ferroelectricity, trentzsch201628nm}. Despite their distinct physical properties, these RAM technologies, i.e.\ MRAMs, OxRAMs, PCRAMs, and FeRAMs, all store memory using resistances.

Due to their fundamentally different physical characteristics, each technology has advantages over the others. State-of-the-art MRAMs, especially utilizing the SOT mechanism, exhibit simultaneously fast switching speed~($\sim$ns) and high endurance (10$^{\textrm{14}}$)~\cite{huai2008spin, carboni2016understanding}. Furthermore, MRAMs are relatively more robust, whereas other memory technologies, especially OxRAMs, can be prone to noise which degrades neural network performance~\cite{majumdar2021model, brivio2022hfo2}. Although FeRAM technology competes with MRAM in terms of performance, particularly in energy efficiency and endurance, it remains a relatively new technology that faces challenges with scalability and device-to-device variability~\cite{banerjee2020challenges, mulaosmanovic2021ferroelectric}. 

Nevertheless, MRAM has its fair share of challenges to overcome~\cite{bhatti2017spintronics}. Since the principal mechanism for resistance switching relies on the tunneling barrier's thickness, precision in the manufacturing process is important. Slight variations in thickness can result in significant performance disparities and affect the device's functionality.
Other limitations of the MRAM technology include a relatively small ON/OFF ratio and the presence of only two states of resistance.

Research in spintronics-based memories is advancing towards mitigating these challenges. Utilizing effects such as SOT has already addressed some of the issues present in STT-MRAM, such as the need for high write currents and longer switching times~\cite{saha2022comparative}. Additionally, memory devices based on magnetic textures are being explored for their immense potential due to their unique physical characteristics. In summary, spintronics-based materials are well-suited for edge applications, and the ongoing research in this field holds promise for realizing low-power, efficient, and fast computations in the future.

\section{Summary and outlook}
\label{sec:6}

\changes{In this chapter, we have demonstrated the significant potential of the interdisciplinary field of Neuromorphic Spintronics to enhance various aspects of modern computing technologies. In section~\ref{sec:1}, we emphasize the suitability of spintronic materials as a foundation for implementing bio-inspired neuromorphic computing. The subsequent sections~\ref{sec:2}, \ref{sec:3}, \ref{sec:4}, and \ref{sec:5} explore different specialized use cases where this concept could have a substantial impact. While these examples vary widely in their applications, they are all unified by their spintronic material-based implementation and underlying neuromorphic principles.} 

Moreover, the field of spintronics is witnessing rapid developments, greatly increasing its potential for neuromorphic applications. One exciting area of progress is the exploration of magnetic textures not only in 2D but also in 3D~\cite{wei2021dzyaloshinsky, vedmedenko2020the}, greatly expanding the range of possibilities. In the third dimension, the complexity of physical processes and responses increases due to interactions with more neighbors, offering improved functionality and tunability. Additionally, 3D systems allow for higher-density processing power and more readout nodes, leading to compact devices and increased error robustness~\cite{everschor2024topological}.

Apart from 3D textures, substantial advancements have been made in the domain of multiferroic materials, where coexisting order parameters can significantly enhance the complexity and functionality of devices~\cite{eerenstein2006multiferroic, everschor2024topological}. By leveraging multiple order parameters, time scales, and length scales, the concept of multi-physics enhances the opportunities for parallel, in-materio, neuromorphic computing. 

As computing devices become increasingly personalized, brain-computer interfaces necessitate the integration of computing elements into the human body. In this context, organic spintronics offers significant advantages due to its cost-effective fabrication, lightweight nature, biocompatibility, and biodegradability~\cite{joshi2016spintronics}. The convergence of these technologies promises to revolutionize how we interact with and utilize computing systems, paving the way for more seamless and integrated solutions. Overall, the advancements discussed throughout this chapter highlight the transformative potential of implementing intelligence with cutting-edge physical systems, paving the way for technological breakthroughs.

\begin{acknowledgement}
We are grateful to Robin Msiska, Maria Azhar, Dennis Meier, Hidekazu Kurebayashi, Jack C. Gartside, and Kerem Camsari for enlightening discussions on various topics, which greatly helped us write the book chapter. We acknowledge funding from the Deutsche Forschungsgemeinschaft (DFG, German Research Foundation) Project-ID 320163632, 403233384 (SPP Skyrmionics), 405553726 – CRC/TRR 270, project B12, 505561633 in the TOROID project co-funded by the French National Research Agency (ANR) under Contract No. ANR-22-CE92-0032, and the Emergent AI Center funded by the Carl-Zeiss-Stiftung.
 \end{acknowledgement}

\bibliographystyle{unsrt}
\bibliography{references.bib}
\end{document}